\documentclass[a4paper,fleqn]{cas-sc}
\usepackage[authoryear]{natbib}

\def\tsc#1{\csdef{#1}{\textsc{\lowercase{#1}}\xspace}}
\tsc{WGM}
\tsc{QE}
\usepackage[nointegrals]{wasysym}
\usepackage{amsmath}
\usepackage{amssymb} % need for \lesssim
\usepackage{graphics}
\usepackage{dcolumn}

\usepackage{lineno}
\usepackage{xcolor,soul}
\sethlcolor{yellow}

\newcommand{\colorindex}[2]{\filter{#1}$\!-$\filter{#2}}
\newcommand{\degsym}{^{\circ}}
\newcommand{\filter}[1]{$\mathit{#1}$}
\newcommand{\noteindex}[1]{\(^{\rm #1}\)}
\newcommand{\polenum}[1]{$P_{_#1}$}

\begin{document}
\let\WriteBookmarks\relax
\def\floatpagepagefraction{1}
\def\textpagefraction{.001}

\shorttitle{Spin vectors in the Koronis family: V. (3032) Evans}

\shortauthors{Slivan et al.}  

\title[mode = title]{Spin vectors in the Koronis family: V. Resolving the ambiguous rotation period of (3032) Evans}

\author[adrMIT12,adrWellesley]{Stephen M. Slivan}[orcid=0000-0003-3291-8708]
\cormark[1]
\cortext[cor1]{Corresponding author}
\ead{slivan@mit.edu}

\author[adrUnion]{Francis P. Wilkin}[orcid=0000-0003-2127-8952]

\author[adrMIT12]{Claire McLellan-Cassivi}

\author[adrMIT12]{Michael J. Person}[orcid=0000-0003-0000-0572]

\affiliation[adrMIT12]{
organization={Department of Earth, Atmospheric, and Planetary Sciences,
              Massachusetts Institute of Technology, Rm. 54-424},
addressline={77 Massachusetts Avenue},
city={Cambridge},
state={MA},
statesep={},
postcode={02139},
country={USA}}

\affiliation[adrWellesley]{
organization={Department of Astronomy, Whitin Observatory, Wellesley College},
addressline={106 Central Street},
city={Wellesley},
state={MA},
statesep={},
postcode={02481},
country={USA}}

\affiliation[adrUnion]{
organization={Department of Physics and Astronomy, Union College},
addressline={807 Union Street},
city={Schenectady},
state={NY},
statesep={},
postcode={12308},
country={USA}}

\begin{abstract}
A sidereal rotation counting approach is demonstrated by resolving an
ambiguity in the synodic rotation period of Koronis family member
(3032)~Evans,
whose rotation lightcurves' features did not easily distinguish between
doubly- and quadruply-periodic.
It confirms that Evans's spin rate does not exceed the rubble-pile
spin barrier
and thus presents no inconsistency with being a $\sim$14-km
reaccumulated object.
The full spin vector solution for Evans is comparable to those for the
known prograde low-obliquity comparably-fast rotators in the Koronis
family,
consistent with having been spun up by YORP thermal radiation torques.
\end{abstract}

% Keywords
\begin{keywords}
asteroids
\sep
asteroids, rotation
\sep
photometry
\end{keywords}

\maketitle

\section{Introduction}
\label{INTR-SEC}

Spin properties studies of asteroid families constrain models of
spin evolution,
avoiding
the difficulties of interpreting properties
of objects which do not share a common origin and dynamical history.
\citet{SLIV08a,SLIV23b} have described
a long-term observing program undertaken to
increase the sample of determined Koronis family spin vectors,
during which (3032) Evans
($H=11.75$, $D$\ $\sim$\ 14~km)
was observed by \citet{SLIV18} as a smaller Koronis member target of
opportunity.

The rotation lightcurve amplitudes of Evans do not exceed 0.2 mag.
\citep{SLIV18,DITT18},
and although the lightcurves are doubly-periodic in about 1.7~h,
that spin rate would
exceed the rubble-pile spin barrier \citep[Fig.~1]{PRAV02}
and require Evans to be an
extraordinary case of a $\sim$14-km solid rock
among the gravitational aggregates comprising the Koronis family.
A rotation period twice as long near 3.4 h is favored instead
which yields quadruply-periodic lightcurves;
\citet{HARR14} have calculated that a lightcurve dominated by a fourth
harmonic can account for the observed amplitude.
The longer period is corroborated by
indications of systematic asymmetry
in the shape of the lightcurves observed
at the 2008 apparition aspect,
but the shape difference between the two halves of the
folded composite lightcurve
is not large enough to be conclusive.

The combination of
the statistically weak distinction of the longer period
given the available data,
and
the science implications if the shorter period were correct,
together
motivate the effort to resolve the ambiguity in this period.
This paper reports additional lightcurve observations,
the identification of the true period,
and spin vector and model shape results for Evans.
In that context it discusses as a case study an approach to resolve
the ambiguous synodic rotation period,
even in the absence of detectable asymmetry in the lightcurves'
shapes,
by analysis of lightcurve epochs from multiple apparitions to
constrain the sidereal rotation period.
Relatively limited details about determining sidereal periods
have been available in the literature
\citep{TAYL83,MAGN86,KAAS01b};
the analysis in this paper
is based mainly on the subsequent
discussion by \citet{SLIV12b,SLIV13}.

\section{Observations}
\label{OBS-SEC}

%%% OBSERVING
Lightcurves of Evans were recorded during three apparitions
using CCD imaging cameras at four different observatories;
the observing circumstances are summarized in
Table~\ref{SUMMARYCIRC-TBL}.
Listed are UT date range of observations,
number of individual lightcurves $N_{\rm lc}$ observed,
approximate J2000 ecliptic longitude $\lambda_{\rm PAB}$
and latitude $\beta_{\rm PAB}$ of the phase angle bisector (PAB),
range of solar phase angles $\alpha$ observed,
and filter(s) and telescope(s) used.
Information about the telescopes, instruments, and observers appears
in Table~\ref{TEL-TBL}.

\begin{table*}[width=.9\textwidth,cols=2,pos=h]
\caption{Observing circumstances summarized by lunation.}
\label{SUMMARYCIRC-TBL}
\begin{tabular*}{\tblwidth}{@{}lrrrrll@{}}
  \toprule
UT date(s) &
\multicolumn{1}{c}{$N_{\rm lc}$} &
\multicolumn{1}{c}{$\lambda_{\rm PAB}$} &
\multicolumn{1}{c}{$\beta_{\rm PAB}$}
&
\multicolumn{1}{c}{$\alpha$} &
Filter(s) &
Telescope(s)

\\ \midrule
2018 Feb 7--15 & 3 & 127$\degsym$ &   +3$\degsym$ &  4$\degsym$--7$\degsym$  & \filter{R}, \filter{r'} & 0.61-m Sawyer \\
2020 Jul 16--28& 7 & 289$\degsym$ & $-$2$\degsym$ &  2$\degsym$--7$\degsym$  & \filter{R}              & 0.36-m C14 \#2 \\
2020 Aug 9--24 & 5 & 290$\degsym$ & $-$2$\degsym$ & 12$\degsym$--16$\degsym$ & \filter{R}, colorless   & 0.36-m C14 \#2,3; 0.61-m CHI-1; 0.50-m CHI-2,4 \\
2020 Oct 2     & 1 & 296$\degsym$ & $-$2$\degsym$ & 21$\degsym$              & colorless               & 0.61-m CHI-1 \\
2023 Jan 13    & 1 & 133$\degsym$ &   +3$\degsym$ &  8$\degsym$              & \filter{g',r',i',z'}    & 1.52-m TCS (Telescopio Carlos S\'{a}nchez) \\
\bottomrule
\end{tabular*}
\end{table*}

\begin{table*}[width=.9\textwidth,cols=2,pos=h]
\caption{Telescopes, instruments, observers.}
\label{TEL-TBL}
\begin{tabular*}{\tblwidth}{@{}llllll@{}}
\toprule
Telescope      & Location                                 &Instrument&Detector  &Image       & Notes \\
               &                                          && field of         & scale &      \\
               &                                          && view ($'$)    & ($''$/pix) &      \\
\midrule
0.61-m Sawyer  & Whitin Obs., Wellesley, MA            &FLI PL23042  &  20 $\times$ 20  & 1.2  & a,e \\
0.36-m C14 \#2 & Wallace Astrophys. Obs., Westford, MA &SBIG STL-1001&  22 $\times$ 22  & 1.3  & b,e \\
0.36-m C14 \#3 & Wallace Astrophys. Obs., Westford, MA &SBIG STL-1001&  20 $\times$ 20  & 1.2  & b,e \\
0.61-m CHI-1   & Telescope Live, El Sauce Obs., Chile  &FLI PL9000   &  32 $\times$ 32  & 1.2  & c,f \\
0.50-m CHI-2   & Telescope Live, El Sauce Obs., Chile  &FLI PL16803  &  66 $\times$ 66  & 0.96 & c,f \\
0.50-m CHI-4   & Telescope Live, El Sauce Obs., Chile  &FLI PL16803  &  66 $\times$ 66  & 0.96 & c,f \\
1.52-m TCS     & Teide Obs., Tenerife, Spain           &MuSCAT2      & 7.4 $\times$ 7.4 & 0.44 & d,e \\
\bottomrule
\end{tabular*}
\begin{flushleft}
Notes:
(a)
Observers
N.\ Gordon, C.\ Miller, A.\ Escamilla Salda\~{n}a,
L.\ Sheraden-Cox, N.\ Tan.
(b)
Observers C.\ McLellan-Cassivi, R.\ Shishido, N.\ Wang.
(c)
Observer F.\ Wilkin.
(d)
Observer H.\ McDonald.
(e)
Observing and data reduction procedures
as previously described by \citet{SLIV08a}.
(f)
Images measured using the AstroImageJ application
\citep{COLL17}.
\end{flushleft}
\end{table*}

\begin{figure*}
\centering
\includegraphics[width=165mm]{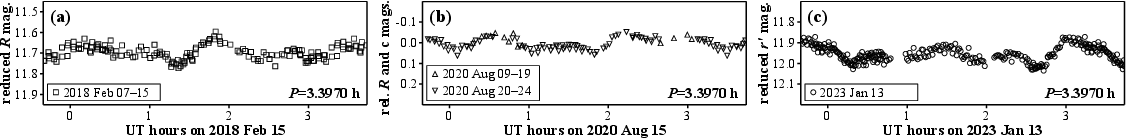}
\caption{Composite lightcurves of (3032) Evans, light-time corrected and
folded at 3.3970 h showing one period plus the earliest and latest
10\% repeated.
Legends give UT dates of observations.
(a) 2018 apparition: The error in the \filter{R} mag.  determined for
the comparison star,
based on observations with Landolt standard star SA097-249
\citep{LAND92},
is 0.007 mag.
(b) 2020 apparition, August lunation.
(c) 2023 apparition: Only the \filter{r'} data are shown.
The brightness zero-point is based on the comparison star's catalog
\filter{r'} magnitude from the APASS DR10 catalog \citep{HEND19} with
error 0.004 mag.}
\label{LCFIGS-FIG}
\end{figure*}

%%% PROCESSING & %%% MEASURING
Images were processed and
measured using standard
techniques for synthetic aperture photometry.
%%% COMPOSITING
Lightcurves were reduced for light-time,
and standard system-calibrated brightnesses were reduced to unit distances.
%%% OUTCOME
One composite lightcurve
selected from each apparition
is shown in Fig.~\ref{LCFIGS-FIG},
where nights of uncalibrated relative photometry
have been shifted in brightness for best fit to
their respective composites.
Color indices determined
from the 2023 data using the MuSCAT2 instrument \citep{NARI19}
which imaged simultaneously using multiple filters are
\colorindex{g'}{r'} = $0.610 \pm 0.025$,
\colorindex{r'}{i'} = $0.187 \pm 0.004$, and
\colorindex{i'}{z'} = $0.058 \pm 0.147$.
No significant color variation was detected
among the three rotationally resolved data sets,
whose sample standard deviations
are 0.024, 0.020, and 0.026 mag.,
respectively.

\section{Resolving the period ambiguity}
\label{RESOLVE-SEC}

\citet{SLIV18} identified
three possible approaches for
additional observations
to identify the correct period:

{\em
Approach A:
Test the previously unobserved viewing geometry
for greater asymmetry in lightcurves.}
Earth-based observing geometries of Evans available during the years
spanned by the lightcurves are clustered around four
ecliptic longitudes roughly 90 degrees apart,
three of which had been previously observed.
The data from 2020 (Fig.~\ref{LCFIGS-FIG}b)
represent the previously unobserved aspect,
but the lightcurve shape asymmetry
is smaller than that seen
in 2008
and thus does not distinguish which period is correct.

\begin{figure}
\centering
\includegraphics[scale=1.00]{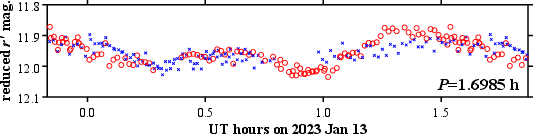}
\caption{The same data as Fig.~{\protect\ref{LCFIGS-FIG}}c but
folded at
1.6985 h for a doubly-periodic lightcurve.
The
graph symbols distinguish in which half of rotation
phase each data point would appear if it were folded at 3.3970 h instead.
The significant difference in the subset lightcurves' shapes between
1.1 h and 1.5 h
in rotation phase
indicates that this shorter period is
not correct.}
\label{LC23-DP-FIG}
\end{figure}

{\em
Approach B:
Lower-noise data near viewing geometry of greatest lightcurve asymmetry.}
The lightcurve from 2023 (Fig.~\ref{LCFIGS-FIG}c)
records a lightcurve similar to that seen in 2008,
and the use of a larger telescope at a dark site
improves on the data quality
and clearly establishes significant asymmetry in the lightcurve.
Folding these data at the 1.7 h period
thus rules out the doubly-periodic alias solution
(Fig.~\ref{LC23-DP-FIG}),
confirming that the correct rotation period is 3.4 h with
quadruply-periodic lightcurves.

{\em Approach C:
Assemble a multi-apparition data set
to determine the sidereal rotation period.}
The opportunity in 2023 to obtain data of high enough quality to
resolve the synodic period ambiguity directly was unexpected.
Even if instead the lightcurves had been of greater noise
comparable to the previous observations,
they still provide a sixth apparition of dedicated lightcurves
for determination of the sidereal rotation period.
This last approach to identifying the correct period
distinguishes that the fractional rotations induced by angular changes
in the direction vector
affect the epochs differently in rotation phase
depending on the rotation period.
The approach is more powerful than the others because it does not
depend on detecting asymmetry in the lightcurve shape,
and motivates documenting the Evans analysis as a case study for reference.

\section{Epochs analysis for sidereal rotations}

The lightcurves reported in Sec.~\ref{OBS-SEC} were combined with
published lightcurves of Evans \citep{SLIV18} for a data set
comprising six apparitions spanning 15 years including all four
available observing geometries,
and satisfying the
data set characteristics
discussed by \citet{SLIV12b,SLIV13}.

% SIEVE ALGORITHM

The sidereal period of Evans is constrained
by the time intervals between repeating lightcurve features,
analyzed using the sieve algorithm
described in detail by \citet{SLIV13}.
The approach
identifies consistent sidereal rotation counts
under the assumption that the epochs defining a given interval correspond to
either the same asterocentric longitude or to reflex longitudes.
It calculates the maximum possible fractional rotations
induced by changes in the direction vector
for both the prograde- and the retrograde-spin cases,
making an approximation
suitable for the small $3\degsym$ orbit inclination of Evans
to calculate direction vector changes
as the differences in ecliptic longitudes,
but otherwise disregarding effects from changing polar aspect angle.

% SCAN

Complementing the sieve analysis for Evans is
an RMS fit error noise spectrum analysis calculated using a model based on
the same underlying assumptions,
noting that the model requires that
all of the epochs be referenced to the same time zero point,
which is more restrictive than the sieve
that depends only on the intervals between pairs of epochs.
The spectrum calculation is a simplification of sidereal photometric
astrometry \citep{DRUM88} limiting to the two equatorial aspect cases
for prograde spin and retrograde spin,
and without determining their corresponding pole ecliptic coordinates.
Given $N$ epochs $t_i$ as time differences from the earliest epoch,
the model is
\begin{linenomath}
\begin{align}
t_i = Pn_i + \tau_0
\end{align}
\end{linenomath}
where the fixed slope $P$ is a trial sidereal period
and $\tau_0$ is the fitted intercept.
The independent variable
$n_i$ is the calculated number of elapsed sidereal rotations
at epoch $t_i$:
\begin{linenomath}
\begin{align}
\label{NROT-EQN}
n_i = 0.5 \mbox{ INT}\left[ 2 \left( \frac{t_i}{P} \mp k_i \right) +0.5 \right] \pm k_i
\end{align}
\end{linenomath}
The maximum fraction of rotation $k_i$ induced by direction vector changes
is calculated
making the same approximation as is used for the sieve,
by
$k_i = \Delta \lambda_i / 360\degsym$
where $\Delta \lambda_i$
is the angular difference of the
PAB longitude for epoch $t_i$
from that for the earliest epoch.
The signs of the $k_i$ in Eq.~\ref{NROT-EQN} depend on whether
calculating for the asteroid spin direction that is the same as (upper
signs) or opposite (lower signs) the orbit direction.
For each trial period
the least-squares solution for the model intercept
is
\begin{linenomath}
\begin{align}
\tau_0 = \frac{1}{N}\left(\sum t_i - P\sum n_i\right)
\end{align}
\end{linenomath}
where the sums are over the $N$ epochs,
and the corresponding RMS error for the spectrum is
\begin{linenomath}
\begin{align}
\sqrt{\frac{1}{N-1} \sum \left (t_i - (P n_i + \tau_0) \right)^2}
\label{RMS-EQN}
\end{align}
\end{linenomath}
calculated for a series of trial sidereal periods over the range of
the synodic period constraint.
The separation of local minima
\citep[Eq. 2]{KAAS01b}
informs the choice of period step size---for visual clarity a step of 
$0.1 \times \left(P_{\rm syn}^2 / 2 T\right)$
samples ten points per local minimum
for synodic period $P_{\rm syn}$ and
time span $T$ between the earliest and
latest epochs.

\subsection{Application to (3032) Evans}
% DETERMINE EVANS EPOCHS AND ERRORS

Table~\ref{EPOCHS-TBL} summarizes the epochs used for the analyses,
one selected per apparition,
with their measurement errors $\sigma(t_i)$ and the
J2000.0 ecliptic longitude $\lambda_{\rm PAB}$ and latitude $\beta_{\rm PAB}$
of the corresponding phase angle bisectors.
Epochs were measured from the composite lightcurves,
in each case locating a maximum of the
fourth harmonic of a Fourier series model
fit to the lightcurves
using the 3.397-h synodic rotation period as the fundamental.
Despite the rather low amplitude,
the lightcurves' pattern of alternating brighter and fainter maxima is
relatively symmetric in time,
making it straightforward to choose an epoch
corresponding to one of the brighter unfiltered maxima
as suitable for analysis of both the doubly- and quadruply-periodic cases.
The epoch measurement errors $\sigma(t_i)$
are based on the RMS error of the corresponding
unfiltered Fourier series fit model to the lightcurve data,
calculating the corresponding error in time by dividing by the model's
steepest slope,
on the brightness increase immediately preceding the brightest maximum.

\begin{table}
\caption{Epochs measured from the lightcurves.}
\label{EPOCHS-TBL}
\begin{tabular*}{\tblwidth}{@{}llrrl@{}}
\toprule
UT date & Epoch (UT h) & \multicolumn{1}{c}{$\lambda_{\rm PAB}$}&\multicolumn{1}{c}{$\beta_{\rm PAB}$} & Ref. \\
\midrule
2008 Jan 16 & $0.34 \pm 0.07$ &  $113.2\degsym$ & $+1.7\degsym$ &  a\\
2009 May 11 & $0.05 \pm 0.07$ &  $192.0\degsym$ & $+3.6\degsym$ &  a\\
2016 Nov 05 & $2.26 \pm 0.10$ &  $ 30.3\degsym$ & $-3.4\degsym$ &  a\\
2018 Feb 15 & $1.85 \pm 0.05$ &  $126.5\degsym$ & $+2.6\degsym$ &  b\\
2020 Aug 15 & $0.55 \pm 0.06$ &  $289.6\degsym$ & $-1.9\degsym$ &  b\\
2023 Jan 13 & $3.14 \pm 0.03$ &  $132.9\degsym$ & $+2.6\degsym$ &  b\\
\bottomrule
\end{tabular*}
\begin{flushleft}
Data references:
(a) \citet{SLIV18}.
(b) this work.
\end{flushleft}
\end{table}

% APPLY SIEVE ALGORITHM

% APPLICATION NOTES - EPOCH ERRS, EPOCH RANGE HALF-WIDTHS, INTERVAL UNC.

To avoid possible dependence of the analysis outcomes on the
atypical lower noise of the 2023 lightcurve,
an increased measurement error of 0.10 h was adopted for the 2023 epoch.
Epoch range half-widths of $2.5\sigma(t_i)$ were used for the sieve,
and in a change from the description in \citet{SLIV13}
the interval ranges were calculated from the epoch
ranges as sums in quadrature.
The synodic period constraint adopted as
the test range of possible sidereal periods was
$P_{\rm syn} \pm 2.5\sigma(P_{\rm syn})$.

% QUADRUPLY PERIODIC CASE

\begin{figure}
\centering
\includegraphics[scale=1.00]{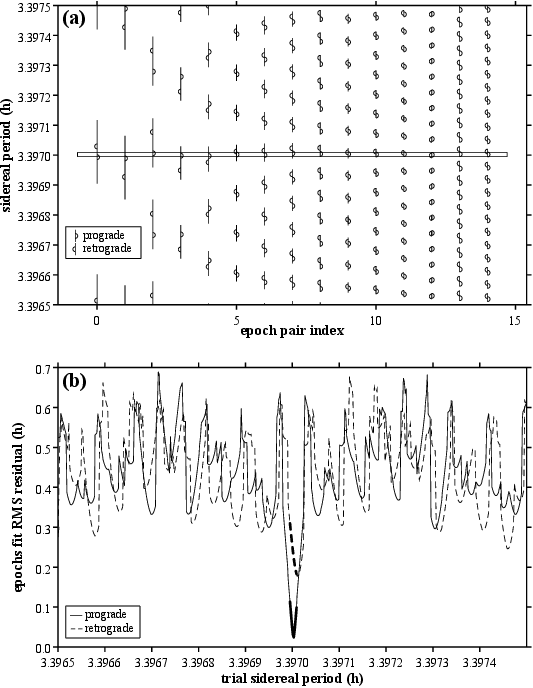}
\caption{(a) Sieve algorithm identifies the sidereal period constraint
for the quadruply-periodic case.
Horizontal axis indices each represent the time interval between a
pair of epochs,
with longer intervals to the right.
Semicircle symbols represent sidereal periods calculated from every
possible number of rotations that could have elapsed during the
interval,
and vertical bars represent period ranges based on epoch range
half-widths of $2.5\sigma(t_i)$.
The thin horizontal rectangle straddling 3.3970 h marks the single
range of periods that is allowed by all fifteen time intervals.
(b) RMS error spectrum for the quadruply-periodic case.
Bold highlighting marks the allowed period range from the sieve
algorithm Fig.~{\protect\ref{SIEVESCAN34-FIG}}a which contains the prograde RMS
minimum at 3.39700 h.}
\label{SIEVESCAN34-FIG}
\end{figure}

Checking the quadruply-periodic case
$P_{\rm syn} = 3.3970 \pm 0.0002$ h \citep{SLIV18}
with the sieve algorithm
identifies an unambiguous sidereal rotation count solution
(Fig.~\ref{SIEVESCAN34-FIG}a) that
is insensitive to alternate choices of epoch range half-widths
down to $1.0\sigma(t_i)$,
and is
corroborated by the noise spectrum (Fig.~\ref{SIEVESCAN34-FIG}b).
The maximum interval between the 2008 and 2023 epochs corresponds to
38689 sidereal rotations, and
the epochs are sufficient to distinguish that the spin is prograde.

% DOUBLY PERIODIC CASE

\begin{figure}
\centering
\includegraphics[scale=1.00]{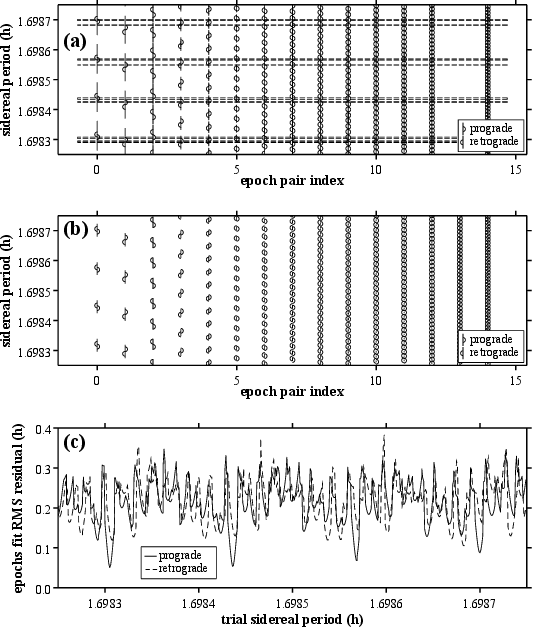}
\caption{Similar to Fig.~{\protect\ref{SIEVESCAN34-FIG}},
but for the doubly-periodic case.
(a) Sieve calculation using epoch range half-widths of $2.5\sigma(t_i)$,
for which the epoch pair at index 13 does not contribute a constraint
and is not plotted.
All of the allowed period ranges are sensitive to the half-widths as
indicated by dashed markings.
(b) Sieve calculation using epoch range half-widths of $1.0\sigma(t_i)$,
finding no allowed period ranges.
(c) The fit error spectrum
does not indicate a secure solution.}
\label{SIEVESCAN17-FIG}
\end{figure}

Repeating the analysis but for the doubly-periodic case
finds only unconvincing solutions whose existence is
sensitive to the choice of epoch range half-widths
(Fig.~{\protect\ref{SIEVESCAN17-FIG}}a);
in fact, testing a reduction from $2.5\sigma(t_i)$ to $1.0\sigma(t_i)$
leaves no solutions at all
(Fig.~{\protect\ref{SIEVESCAN17-FIG}}b).
The noise spectrum
has several indistinguishable local minima
which also is not characteristic of correct solutions
(Fig.~{\protect\ref{SIEVESCAN17-FIG}}c).
The absence of a secure sidereal rotation count solution for 1.7 h,
and the existence of the secure solution for 3.4 h,
indicates that the 1.7 h period is an alias.

\section{Analysis for spin vector and convex model}
\label{SV-ANALYSES-SEC}

\begin{figure}
\centering
\includegraphics[scale=1.00]{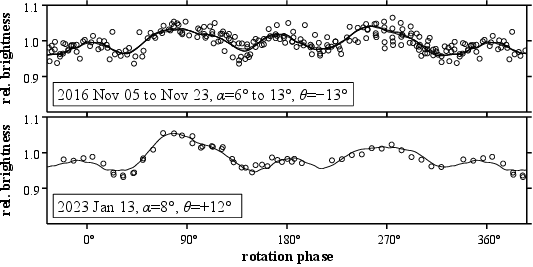}
\caption{Selected model lightcurve fits
with a pole \polenum{2} at (354$\degsym$;$+$70$\degsym$)
as brightness vs. sidereal rotation phase.
Legends give UT dates of the observations,
solar phase angles $\alpha$,
and sub-PAB latitudes $\theta$.
The RMS error of the fit to the entire set of lightcurves
corresponds to 0.021 mag.}
\label{A3032-CIRESID-FIG}
\end{figure}

\begin{figure}
\centering
\includegraphics[scale=1.00]{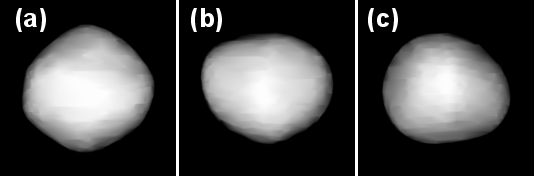}
\caption{Renderings of convex model for pole \polenum{2}.
(a) Polar aspect
showing the four-sided profile
responsible for the quadruply-periodic lightcurves.
(b,c) Equatorial aspects for
the brighter and fainter pairs of lightcurve maxima, respectively.}
\label{SHAPES-FIG}
\end{figure}

Having already resolved the true period from the alias,
the remaining stages of spin vector and convex model analysis
were carried out as described by \citet{SLIV23b}.
Comparably-weighting the lightcurve data by apparitions
produced a lopsided distribution in aspect coverage;
to aid the final convex inversion analyses
the weighting was based on viewing aspects instead.
Selected lightcurve fits are shown in
Fig.~\ref{A3032-CIRESID-FIG}.
Spin vector results
are summarized in Table~\ref{ALL-SV-RESULTS-TBL}:
% SIDEREAL PERIOD
the derived sidereal period $P_{\rm sid}$ and its error,
% POLE SOLUTIONS
the symmetric pair of pole solutions'
J2000 ecliptic
longitudes $\lambda_0$ and latitudes $\beta_0$,
with their respective estimated errors
$\sigma(\lambda_0)$ and $\sigma(\beta_0)$
in degrees of arc,
the corresponding spin obliquities $\varepsilon$,
% MODEL SHAPE
and model shape axial ratios.
The pole solutions satisfy the expected symmetry
with respect to the ``photometric great circle''
\protect\citep[Appendix~A]{MAGN89,SLIV23b},
but the
$3\degsym$
orbit inclination of Evans is small enough to
prevent distinguishing which is the true pole.
The convex model shapes for the two symmetric poles
are essentially mirror images of each other,
to within their poorly-constrained scale factors
in the direction along the polar axis.
Renderings of one model are shown in
Fig.~\ref{SHAPES-FIG},
resembling
a very rounded tetrahedral shape with its spin axis
intersecting opposite edges.

\begin{table}
\caption{Spin vector results.}
\label{ALL-SV-RESULTS-TBL}
\begin{tabular*}{\tblwidth}{@{}rrrrrr@{}}
\toprule
sidereal period $P_{\rm sid}$:&\multicolumn{5}{l}{3.397003 $\pm$ 0.000002 h}\\[0.25pc]
                             &\multicolumn{1}{c}{$\lambda_0$}&\multicolumn{1}{c}{$\sigma(\lambda_0)$}&\multicolumn{1}{c}{$\beta_0$}&\multicolumn{1}{c}{$\sigma(\beta_0)$}&\multicolumn{1}{c}{$\varepsilon$}\\ \cline{2-6}
      spin poles \polenum{1}:&186$\degsym$&  5              & $+$75$\degsym$ &  5          & 18$\degsym$\\
                 \polenum{2}:&354$\degsym$&  5              & $+$70$\degsym$ &  5          & 17$\degsym$\\[0.25pc]
          model axial ratios:&\multicolumn{2}{l}{$a/b$: 1.0\noteindex{a}}&\multicolumn{2}{l}{$b/c$: 1.1\noteindex{a}}&\\
\bottomrule
\end{tabular*}
\begin{flushleft}
Note (a) The axial ratios are very coarse estimates
with uncertainties of at least $\pm 0.1$.
\end{flushleft}
\end{table}

\section{Discussion and Conclusion}
\label{DISC-SEC}

% SIDEREAL PERIOD RESOLUTION

Observations and analysis reported in this work have resolved the
factor of two ambiguity in the synodic rotation period of Koronis
member (3032)~Evans,
confirming that its lightcurves are quadruply-periodic.
The correct period was identified in two ways:
first serendipitously by detecting asymmetry in an atypically
high-quality lightcurve,
and then deliberately by constraining the sidereal rotation period
using the sieve algorithm of \citet{SLIV13} to analyze
a suitable multi-apparition lightcurve data set.
The latter approach can indicate the true period even for objects that
do not show detectable asymmetry in their lightcurves' shapes.

With its 3.4-h rotation period,
Evans's spin rate does not exceed the rubble-pile spin barrier
\citep[Fig.~1]{PRAV02}
and thus presents no inconsistency with being a $\sim$14-km
reaccumulated object member of the Koronis family.

% SPIN VECTOR IN CONTEXT

\citet{SLIV23b} have discussed the Koronis member spin vector sample
completed to $H$\ $\sim$\ 11.3,
in which the smallest prograde-spinning objects also are
the fastest prograde rotators and have low spin obliquities,
consistent with having been spun up by YORP thermal radiation torques
acting more quickly than on larger bodies \citep{RUBI00}.
The spin vector properties of Evans are comparable to
these spun-up prograde objects,
noting that it is both smaller than and faster-rotating than they are.
 
\section*{Acknowledgments}

We thank the
Corps of Loyal Observers, Wellesley Division (CLOWD)
who recorded data at Whitin Observatory:
Naomi Gordon, Cassie Miller, Alejandra Escamilla Salda\~{n}a,
Leafia Sheraden-Cox, and Nicole Tan.
At the Wallace Observatory
we thank
Timothy Brothers
for observer instruction and support,
and summer student observers
Rila Shishido and Nieky Wang.
From the 2023 MIT Astronomy Field Camp
we thank
Helena McDonald for observing at Teide Observatory.
Finlay MacDonald at Union College
assisted with analysis of the 2020 data
from El Sauce Observatory.

Student service observers at Whitin Observatory were supported in part
by grants from the Massachusetts Space Grant Consortium.
The student observers at Wallace Observatory were supported by a grant
from MIT's Undergraduate Research Opportunities Program.
Coauthor F.\ Wilkin received funding from a grant from the Cohen
family.
This article includes observations made at the
Telescopio Carlos S\'{a}nchez (TCS),
operated on the island of Tenerife
by the Instituto de Astrof\'{\i}sica de Canarias
at the Spanish Observatorio del Teide,
utilizing the MuSCAT2 instrument developed by ABC.

\section*{Data Availability}

Datasets related to this article can be found at
http://\allowbreak smass.mit.edu/slivan/lcdata.html.

\bibliographystyle{cas-model2-names}
%\bibliography{refs23}

\end{document}